\documentclass[%
 preprint,
 aps,
pra,
]{revtex4-1}

\usepackage{graphicx}
\usepackage{amsmath}
\usepackage[justification=centerlast]{caption}
\begin{document}

\title{Enhanced topological phase and spin Hall shifts in an optical trap}


\author{Basudev Roy$^1$}
\author{Nirmalya Ghosh$^1$} \email{nghosh@iiserkol.ac.in}
\author{Ayan Banerjee$^1$} \email{ayan@iiserkol.ac.in}
\affiliation{$^1${Department of Physical Sciences, IISER-Kolkata, Mohanpur 741252, India}}
\author{Subhasish Dutta Gupta$^2$} \email{sdghyderabad@gmail.com}
\affiliation{$^2$Department of Physical Sciences, Hyderabad Central University}
\author{Soumyajit Roy$^3$}  \email{s.roy@iiserkol.ac.in}
\affiliation{$^3${EFAML, Materials Science Centre, Department of Chemical Sciences, IISER-Kolkata, Mohanpur 741252, India}}
\date{\today}
\begin{abstract}
The spin orbit interaction (SOI) of light has been in the focus in recent times because of fundamental consequences and potential applications in diverse systems ranging from inhomogeneous anisotropic media to engineered plasmonics and metamaterial strutures. Here we demonstrate perhaps one of the simplest means to realize SOI and the Spin Hall Shift (SHS) using a standard Gaussian TEM$_{00}$ beam in an optical trap. Our system exploits the versatility and interference generated in a stratified medium to control and manipulate SOI and transfer the resulting angular momentum to optically trapped micro-particles. We show that even such a simple setup can lead to an order of magnitude enhancement in the SHS compared to the sub-wavelength shifts typically obtained. Importantly, this leads to controlled rotation of mesoscopic particles using a fundamental Gaussian beam lacking any intrinsic angular momentum.

\begin{description}
\item[PACS numbers: 42.50.Tx, 42.25.Ja, 87.80.Cc, 42.50.Wk]
\end{description}

\end{abstract}

\maketitle
\noindent 
Spin orbit interaction (SOI) -  which couples the spin and orbital degrees of freedom of massive as well as massless particles - can be associated with several fundamental consequences in physics including the fine and hyperfine structure in atoms \cite{cohentann}, and spin Hall effect in electrons \cite{hir99}. In the case of light, SOI and the accompanying geometric phases cause an interdependence of trajectory and polarization (spin). Thus, a change in the  topological degrees of freedom affects the polarization state of light and vice versa. The latter leads to the well-known spin Hall effect (SHE) of light \cite{ono04, bli08, berr05, zhao07, zeev07, haef09, vuo10, her10}, while the former can lead to interconversion between spin angular momentum (SAM) and orbital angular momentum (OAM) of photons, polarization controlled vortices and other different intriguing effects of geometric  phases in scattering and tight focusing \cite{blio09, blio08, maru06, vuo10}. In addition, experimental measurements of effects induced by SOI of light have helped in understanding quantum and condensed matter systems having similar underlying physics. While a few practical applications such as nano-displacement probes \cite{her10}, or generation of optical vortices \cite{brass09} have been developed, in most cases the effects of SOI have been rather small - for example, trajectory shifts reported due to the SHE of light are typically of sub-wavelength magnitudes \cite{zeev07, her10}. 

Optical tweezers, which employ a tightly focused beam to trap and manipulate mesoscopic particles \cite{nated11, dhola11, faz11, mil11, moha11}, have interesting consequences for SOI since tight focusing couples the trajectory and polarization of the propagating light. However, the only report of SOI in optical tweezers in literature is found in Ref.~\cite{zhao07} where SOI mediated angular momentum conservation was shown to affect the angular velocity of rotating particles  for higher order Gaussian beams.  In this paper, we show that the strength of SOI can be significantly magnified in a stratified medium used in the light path of the optical tweezers system. The enhanced SOI gives rise to a large anisotropic linear diattenuation effect (differential attenuation of orthogonal polarizations) so that, for a particular choice of the layers of the stratified medium, a tightly focused linearly polarized Gaussian beam loses its azimuthal symmetry in the focal plane itself with side lobes being formed in the direction of polarization. These side lobes get stronger as the beam propagates, so that at the region of caustics, the beam splits into two with a definite azimuthal pattern. Further spatial evolution in the axial direction recovers the original Gaussian structure.  Such polarization-dependent intensity distribution can be understood as a direct manifestation of the spin redirectional toplogical phase \cite{blio08}. In addition, distinct regions of opposite ellipticity are produced at the edges of the intensity lobes as a manifestation of giant SHE of light.  We perform a rigorous 3d analysis of polarization evolution to calculate the spin Hall shifts (SHS) near the focal plane, and exploit the giant SHE to experimentally demonstrate controllable rotation of asymmetric microparticles in the trap using just a fundamental Gaussian beam with no intrinsic angular momentum. 
\section{Results}
\subsection{Theory and Simulations}
It is now well understood that tight focusing of light leads to a longitudinal component which can be appropriately described in the framework of the Debye-Wolf theory \cite{born, rich59, blio11, her10}, which makes use of the plane wave (spatial harmonic) decomposition of the Gaussian beam. The evolution of each spatial harmonic  can be represented by means of geometric rotations in the azimuthal ($\phi$) and polar ($\theta$) directions with respect to the laboratory frame. A transfer function  $A =  R_z(-\phi)R_y(\theta)TR_z(\phi)$, where $R_i(\alpha),~i=x, y, z$  represents the SO(3) rotation matrix around the $i$ axis by an angle $\alpha$ (for details see Supplementary Information A). The effects of the stratified medium for different polarizations of light, are incorporated through the complex Fresnel transmission and reflection coefficients $T_s~(R_s)$ and $T_p~(R_p)$, respectively. Note that, due to low index contrast, the total field in the stratified medium is dominated by the forward propagating  waves. The final field, after integrating over $\theta$ and $\phi$, is given by
\begin{eqnarray}\label{polarint}
\vec{E}(\rho,\psi,z)&=&  i\frac{kfe^{-ikf}}{2\pi}\int_0^{\theta_{max}}\int_0^{2\pi}
\vec{E}_{res}(\theta,\phi)e^{ikz\cos\theta}\nonumber\\& \times & e^{ik\rho\sin\theta\cos(\phi-\psi)}
sin(\theta)\>d\theta d\phi
\end{eqnarray}
where $r$ is set to $f$ -- the focal length of the focusing lens, and the limit for the $\theta$ integral is set by the numerical aperture of the microscope objective of the optical trap. Using Eqn.~\ref{polarint}, the electric field inside the medium can be written as 
\begin{eqnarray}\label{fieldout}
\left[
\begin{array}{c}
 {E_x^o}\\
 {E_y^o}\\
 {E_z^o}\\
\end{array}
\right ] &= C \left[
\begin{array}{lll}
 I_0 + I_2\cos 2\psi  & I_2\sin 2\psi & 2i I_1\cos\psi \\
 I_2\sin 2\psi & I_0 - I_2\cos 2\psi & 2iI_1\sin\psi \\
 -2iI_1\cos\psi & -2i\sin\psi & I_0+I_2 \\
\end{array}
\right ]& \left[
\begin{array}{c}
 {E_x^i}\\
 {E_y^i}\\
 {E_z^i}\\
\end{array}
\right ]
 \end{eqnarray}
where, the superscripts $i$ and $o$ denote the input and output fields respectively, $I_0, ~I_1$ and $I_2$ being the well-known diffraction integrals (Supplementary Material A), and $C$ is a constant. The effect of SOI is strongly manifested in the case of input linearly $x$-polarized light represented by a Jones vector $\left[\ 1~0~0\ \right]^T$. Since linear polarization can in general  be written as the sum of orthogonal circular polarizations, the output field may be written as
\begin{equation}
\label{jonesveclinpol}
E = I_0\left[
\begin{array}{c}
\ {1}\\
\ {0}\\
\ {0}\\
\end{array}
\right ]
+I_2exp(i2\psi)
\left[
\begin{array}{c}
\ {1}\\
\ {-i}\\
\ {0}\\\end{array}
\right ] + I_2exp(-i2\psi)
\left[
\begin{array}{c}
\ {1}\\
\ {i}\\
\ {0}\\\end{array}
\right ]-2iI_1exp(i\psi)
\left[
\begin{array}{c}
\ {0}\\
\ {0}\\
\ {1}\\\end{array}
\right ] - 2iI_1exp(-i\psi)
\left[
\begin{array}{c}
\ {0}\\
\ {0}\\
\ {1}\\\end{array}
\right ]
\end{equation}
Thus, it is clear that for linearly polarized light, one would obtain right and left circularly polarized components having orbital angular momentum  $l =-2$ and $l = 2$, respectively, as well as linearly polarized longitudinal components with topological charge $l = \pm 1$, with each component satisfying total angular momentum conservation. The associated coefficients $I_0$, $I_2$ and $I_1$ of the transverse (1st, 2nd, and 3rd terms) and the longitudinal (4th and 5th terms) field components determine the strength of the spin orbit angular momentum conversion. The generation of such circular polarization states from a linear input is a manifestation of the SHE. We will subsequently show that the effect of the stratified medium is to cause a giant SHE - or a transverse spatial shift of these individual polarization components so that they would have definite effects on trapped micro-particles. It is also evident from Eqs.~\ref{fieldout} that the total intensity for incident linearly polarized light would be given by \begin{equation}
I(\rho) = \ \left|I_0\right|{^2} + \left|I_2\right|{^2}  \pm  2 {\bf Re}(I_0I_2^{\star})\cos 2\psi 
+ 2 \left|I_1\right|^2 (1 \pm \cos 2 \psi)
\label{intensitylinpolar}
\end{equation}
Here, we introduce a quantity $\mathfrak{D}$ where ${\mathfrak{D} = (\bf Re}\ (I_0I_2^{\star})+ \left|I_1\right|^2)$. $\mathfrak{D}$ is known as the linear diattenuation parameter and is a measure of the polarization dependence of intensity, with the quantity $\mathfrak{D}(1 \pm \cos 2 \psi)$ giving  the  intensity distribution as a function of the input polarization angle $\psi$. Note that $\mathfrak{D}(1 \pm \cos 2 \psi)$ is essentially a a topological phase term that is picked up due to the SOI of the tightly focused light as it propagates through the stratified medium and its magnitude can be used to quantify the strength of SOI in the system. 

It is important to note that the SHS originates due to the finite longitudinal component of the field which leads to transverse energy flows. The separation of the spins can be understood as the shift of the center of gravity of the beam for the two opposite circular polarization states.  Thus, the SHS can be determined from the longitudinal ($P_z$) and the transverse ($P_{x,y}$) components of the Poynting vector, so that a 3d treatment of polarization is warranted. We therefore proceed to build the 3d coherency matrix of the system  (see Supplementary Information A for details) \cite{set02}, from which we define the $9\times 1$ Stokes vectors $\Lambda_{i, i=1, 9}$ to include the longitudinal polarization component as well . Then, the total degree of polarization may be defined as 
\begin{equation}
{\rm DP} = \dfrac{1}{3}\dfrac{\Sigma \Lambda_i^2}{\Lambda_o^2},
\label{dop}
\end{equation}
so that the degree of linear polarization (DLP) and that of circular polarization (DCP) can be obtained from the corresponding linear and circular descriptor Stoke's parameters. The helicity of the polarization is not apparent from the DP, and thus it is necessary to find out the 2-d degree of circular polarization DCP from the 2-d Stokes vectors which are obtained by neglecting the $z$-components of electric field in the 3d coherency matrix. This gives us the
\begin{equation}
{\rm DOCP}_2 = \dfrac{V}{\sqrt{Q^2 + U^2 +V^2}},
\label{docp}
\end{equation}
$Q,~U,~{\rm and}~V$ being the well-known 2-d Stokes vector elements (see Supplementary Information A). Finally, the SHS for circularly polarized light is given by 
\begin{equation}
\Delta y = \pm \dfrac{\Sigma y (P_x + P_y)}{\Sigma P_i},
\label{shs}
\end{equation} with the positive and negative signs corresponding to left and right circularly polarized light respectively, and $i=x,y,z$. Now, for linearly polarized tightly focused light, the constituent circular polarization components evolve in different trajectories resulting in a net spatial shift of the individual components to create spatially separated regions of opposite circular polarization near the focal plane. We go on to show that this spatial separation or SHS is considerably amplified in the presence of a stratified medium leading to controlled rotation of asymmetric micro-particles.

\begin{figure}[h!t!]
{\includegraphics[scale=0.65]{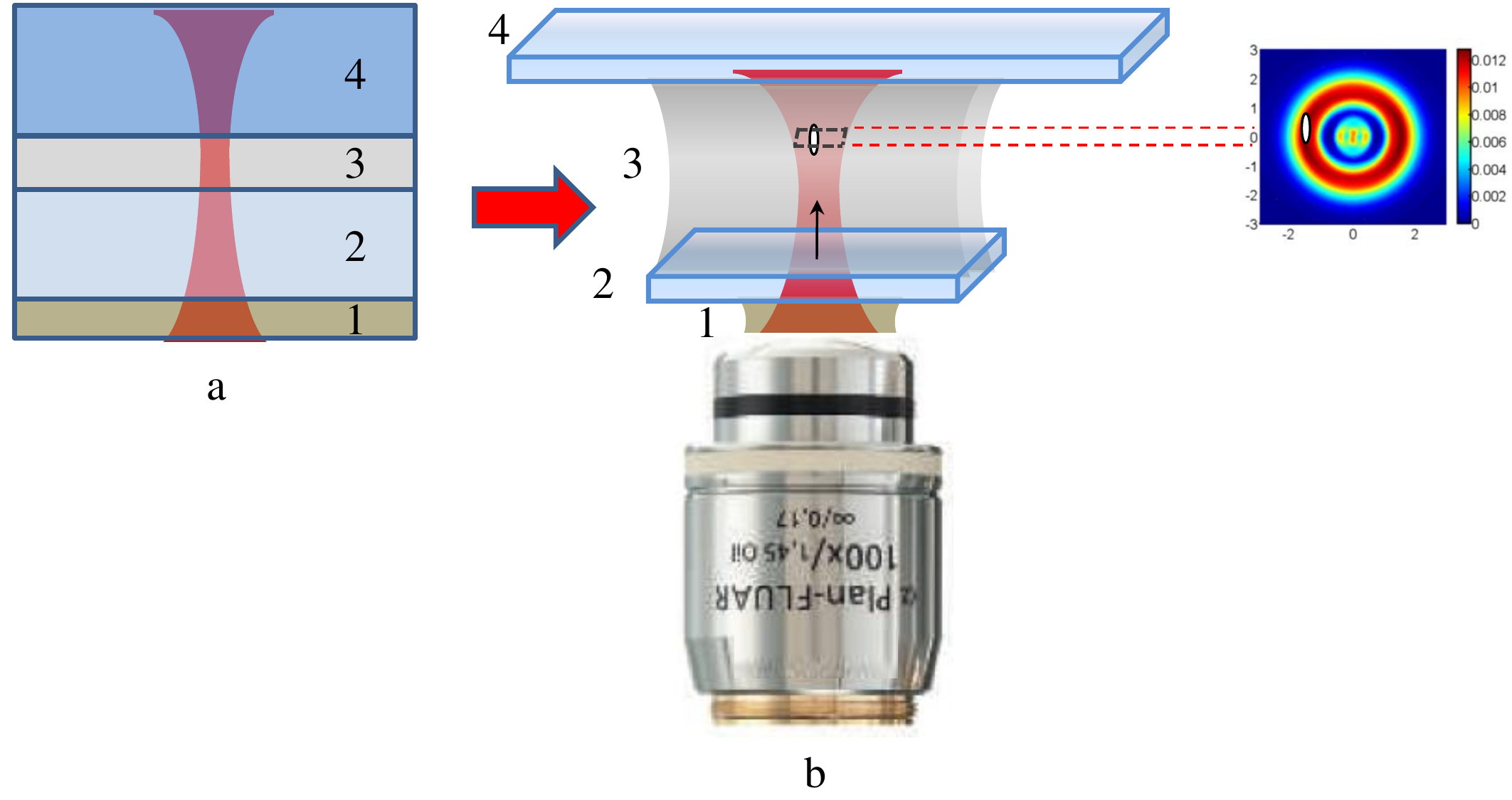}}
\captionsetup{format=plain}
\caption[]{(Color online)(a) Stratified medium (not to scale) formed in our experimental system with a Gaussian beam (direction shown) focused inside the sample solution. The various layers are (1) Objective immersion oil (RI 1.516), (2) Cover slip (RI 1.575), ( aqueous solution of micro-particles (RI 1.33), (4) Top glass slide (RI 1.516). For conventional tweezers, the RI of (1) and (2) are matched. (b) A cartoon of the actual experimental setup with the different  layers of the stratified medium shown physically. The radial distribution of the electric field intensity near the focus (region shown using red dashed lines) is also shown. The intensity distribution is modified from a Gaussian structure into a ring like pattern due to SOI.}
\label{exptsys}
\end{figure}
Using the above theoretical approach, we performed simulations for a stratified medium consisting of four layers that matched our experimental conditions. A schematic of our system is shown in Fig.~\ref{exptsys}(a). The optical tweezers is built around an inverted microscope (Zeiss Axiovert.A1) with a high NA objective (NA=1.4) that tightly focuses the beam into the sample chamber as is shown in Fig.~\ref{exptsys}(b). The sample system consists of the following layers: 1) immersion oil, 2) cover slip, 3) sample aqueous solution, 4) glass slide. For the simulation, we have chosen two different RI values of layer 2 (cover slip). Simulations were carried out for two specific cases, namely, (a) 
{\it a perfectly polarizing cover slip ($T_p = 0,~ T_s \neq 0$, RI = 1.575)}, which would lead to complete conversion of spin to orbital angular momentum, and (b) {\it partially polarizing cover slips ($T_p,~ T_s \neq 0$) of different RI}.

\begin{figure}[h!t!]
 \begin{center}
{\includegraphics[scale=0.6]{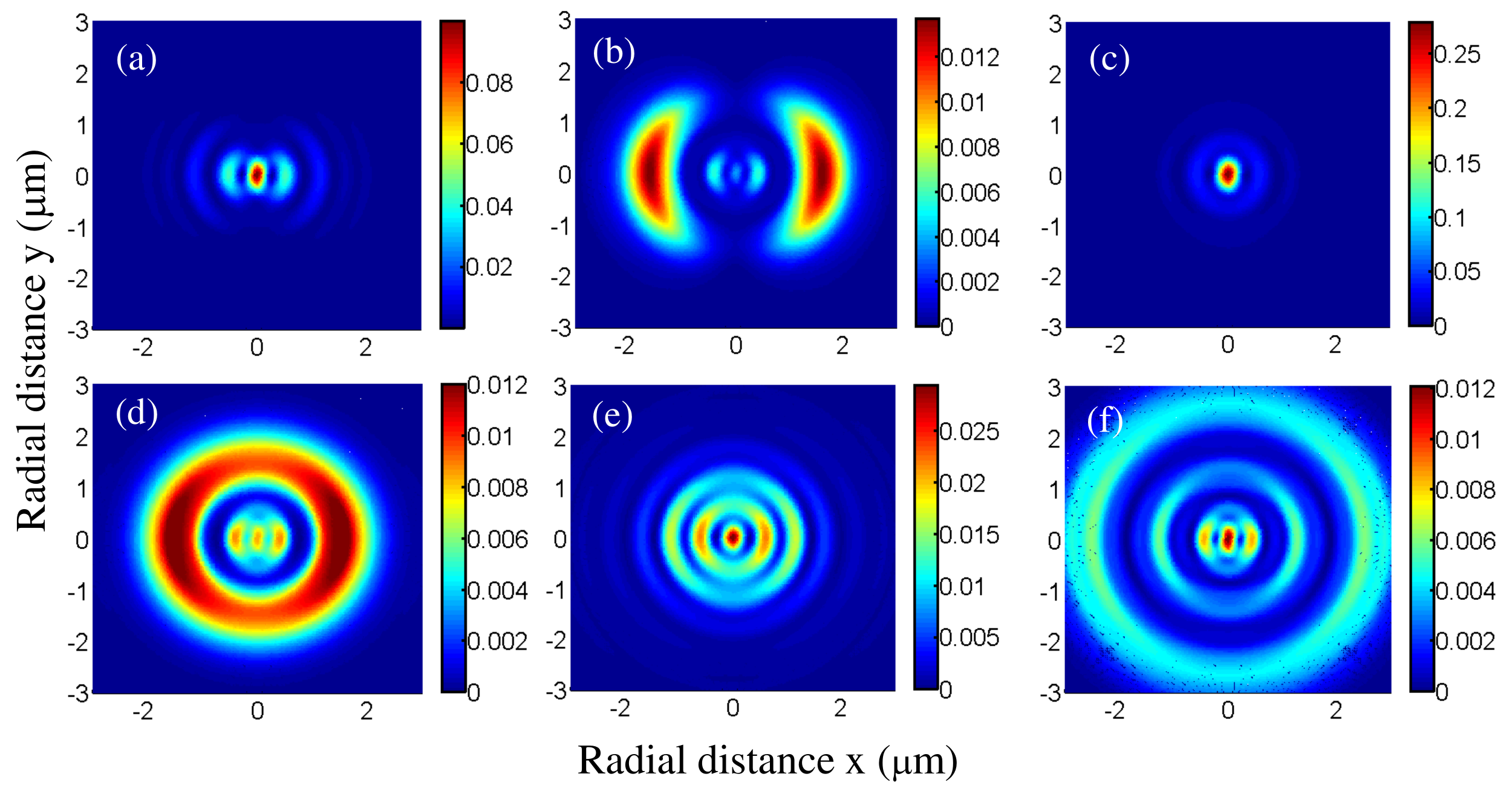}}
 \end{center}
\caption[]{(Color online) Intensity distributions inside sample chamber (layer 3 of Fig.~\ref{exptsys}) of trapping system for different RI and polarization properties of the cover slip (layer 2) for an $x$-polarized input beam. (a) Intensity distribution for polarized cover slip ($T_p = 0, ~T_s \neq 0$) with RI of 1.575, at the focus ($z=0$). Intensity side lobes separated radially by around 2 $\mu$m with strength about 40\% of that of the central lobe are seen. (b) Intensity distribution for same cover slip as in (a) at an axial distance 2 $\mu$m away from the focus ($z=2$). The maximum intensity is now concentrated in two discrete side lobes formed opposite to each other in the direction of polarization of the input beam, and separated radially by around 4 $\mu$m. (c) Intensity distribution at the focus for unpolarized($T_p \neq 0, ~T_s \neq 0$)  cover slip of RI 1.575 at the focus. Side lobes are absent.(d) Intensity distribution of same cover slip as in (c) at $z=2~ \mu m$. The side lobes in the polarization direction are present as in (b), but in the background of a continuous intensity ring of diameter about 4 $\mu$m around the focus. (e) Intensity distribution for cover slip of RI 1.65 at $z=2~ \mu m$. Multiple rings are now visible within the diameter of 3 $\mu$m having higher intensity compared to (d). This implies a higher axial trapping depth as well as larger ring diameters for trapping of particles. (f). Intensity distribution for same cover slip of RI as in (e) at $z=3~ \mu m$. The ring diameter is now around 5 $\mu$m with the intensity levels similar to (d).}
\label{diatt}
\end{figure}

\begin{figure}[h!t!]
 \begin{center}
{\includegraphics[scale=0.6]{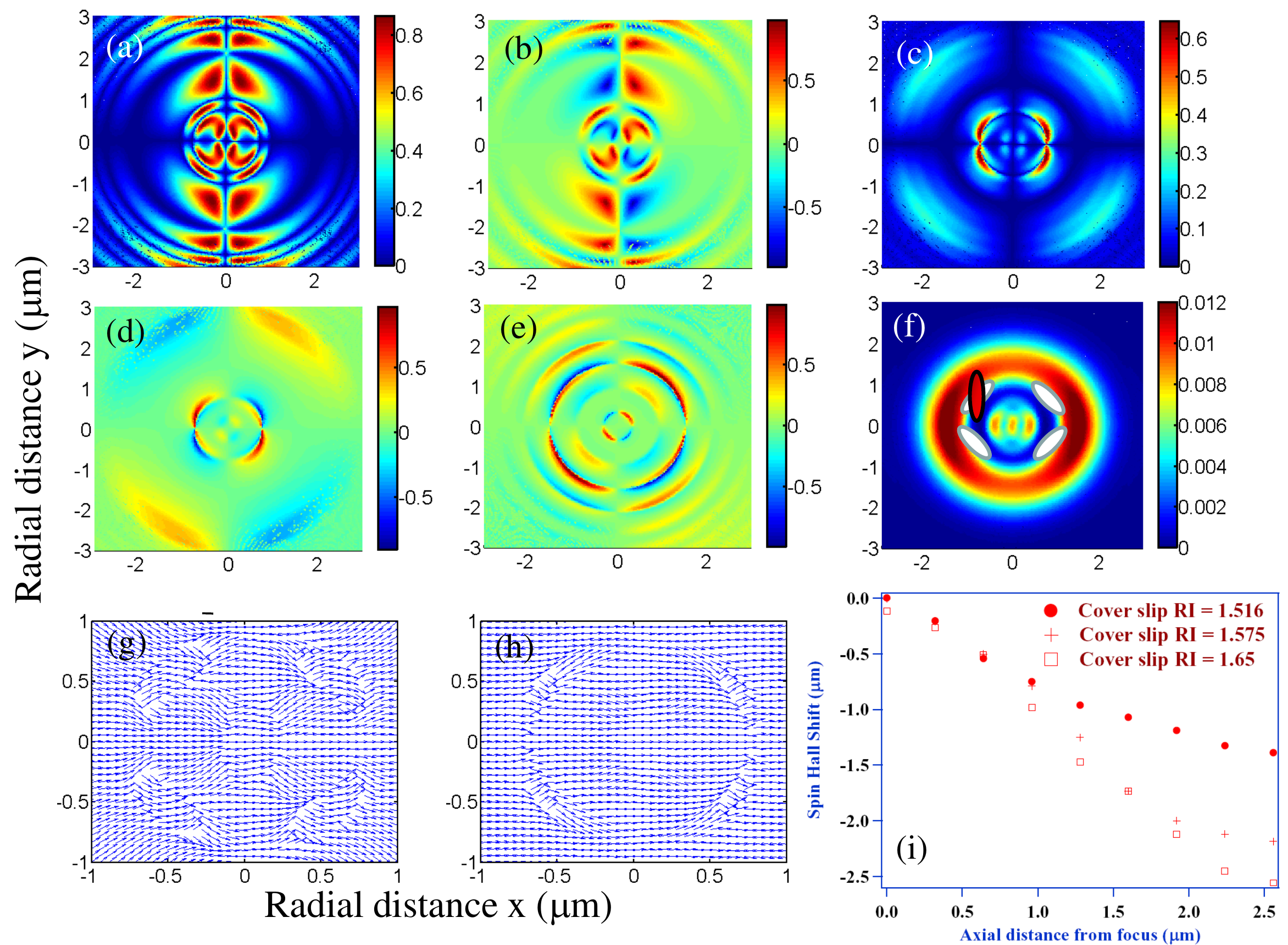}}
 \end{center}
\caption[]{(Color online) Study of the polarization distributions inside sample chamber (layer 3 of Fig.~\ref{exptsys}) of trapping system for different RI and polarization properties of the cover slip (layer 2) for an $x$-polarized input beam. All data are for an axial plane 2 $\mu$m away from the focus. (a) DP for polarized cover slip ($T_p = 0, ~T_s \neq 0$) with RI of 1.575. (b) DCP for (a). The helicities of the polarization lobes are now apparent.(c) DP for unpolarized cover slip with RI of 1.575. (d) DCP for (c).  (e) DCP for RI = 1.65.  (f) Overlay of  DCP regions (white ovals) observed in (d) with the intensity distribution for the same configuration (Fig.~\ref{diatt}(d)). An asymmetric particle (red oval) with form birefringence trapped in the intensity ring and also sampling one of the elliptical polarization regions would be likely to spin. (g) Quiver plot showing the azimuth of the electric field vector for case (b). There are four regions where the electric field describes a vortex, interspersed with these are regions of polarization singularity where the electric field does not exist. (h). Quiver plot showing the azimuth of the electric field vector for case (d). Vortex formation and polarization singularities are seen again, albeit over smaller regions compared to (g). (i) Plots of SHS against axial distance from the focus for layer 2 RIs = 1.516 (RI matched case with only one effective layer in the stratified medium), 1.575, and 1.65.}
\label{elliptfig}
\end{figure}

It is observed that the enhanced SOI breaks down the azimuthal symmetry in the intensity profile so that side lobes are formed even at the focus of the beam for a perfectly polarizing cover slip (Fig.~\ref{diatt}(a)). These become stronger as the beam propagates axially so that a doughnut profile is formed from an input Gaussian intensity distribution (Fig.~\ref{diatt}(b)). For partially polarized cover slips (Figs.~\ref{diatt}(c) and (d)), the side lobes appear in the background of an intensity ring  with the number of rings increasing with RI of the cover slip ((Fig.~\ref{diatt}(e) and (f)). Most importantly, the enhanced SOI results in a large SHS manifested as spatially separated regions having opposite circular polarization at the edges of the intensity lobes (Figs.~\ref{elliptfig}(a)-(e)). A particle trapped at such a region thus samples the high circular polarization as shown in Fig.~\ref{elliptfig}(f). Moreover, tight focusing of input linearly polarized light also leads to generation of phase vortices which in turn create polarization singularities close to the focus (Eq.~\ref{jonesveclinpol}) as shown in Fig.~\ref{elliptfig}(g) and (h). The azimuth of the electric field  vector is calculated from $\gamma = \dfrac{1}{2} \arctan \left(\dfrac{U}{Q} \right),~ U~{\rm and}~Q$ being the well-known Stokes parameters. The magnitude of SHS as a function of axial distance also shows a definite tendency to increase with increasing RI of the cover slip as observed from Fig.~\ref{elliptfig}(h)).

\subsection{Experimental Results}  
\begin{figure}[h!t!]
 \begin{center}
{\includegraphics[scale=0.5]{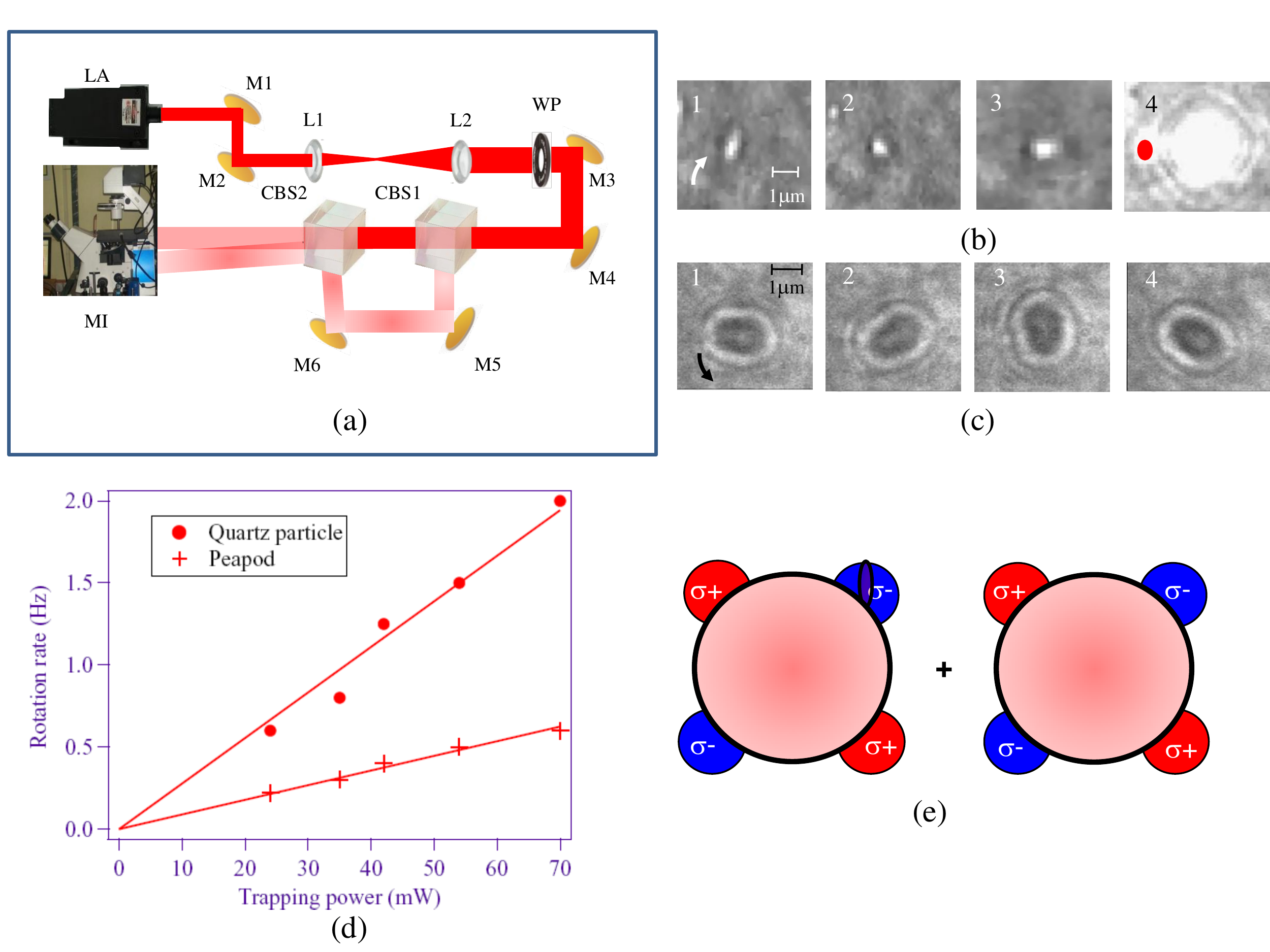}}
 \end{center}
\caption[]{(Color online) (a) Schematic of experimental setup -- MI: Inverted microscope; LA: Solid state laser diode system at 1064 nm; M1-M6: Mirrors; L1, L2: Plano-convex lenses for beam size adjustment to overfill microscope back aperture; WP: Linear (half-wave) wave-plate or retarder; CBS1, CBS2: 50-50 beam splitter cubes for 1064 nm. (b) Time series of a single trapped peapod rotating in the clockwise direction due to elliptical polarization created near the off-axis intensity side lobes (for video see Supplementary Information B). For trace 4, the IR filter in front of the camera has been removed and the particle is observed being trapped in the off-axis intensity ring. (c) Time series of a singe trapped ellipsoid shaped quartz particle rotating anti-clockwise due to the same effect (for video see Supplementary Information B). (d) Rotation speed versus trapping laser power for a representative peapod (red crosses) and quartz particle (red open circles). (e) Cartoon depicting the experimental design for control of rotation of a spinning particle. The particle (black transparent oval) spins in one of the ellipticity regions ($\sigma-$ in this case) produced due to SHS in the trapping light. A second beam of the same input polarization is introduced adjacent to the first beam so that the ellipticity region that is nearest to the particle has opposite helicity of polarization. When the two spots are brought close together the torque on the particle cancels out if the intensities of the two beams are the same so that the particle stops rotating, or reverses its direction of rotation if the intensity of the second beam is higher than the first producing higher opposite torque.} 
\label{exptdata}
\end{figure}
We experimentally demonstrate the results of the effect of such large SOI and SHS on the mechanical motion of microparticles by a number of controlled experiments. A schematic of our setup is shown in Fig.~\ref{exptdata}(a). Images of the intensity distribution near the focal region confirm the results shown in Fig.~\ref{diatt}(d) with a ring-like transverse pattern, which can lead to trapping of particles \cite{hal12}, and their movement along it \cite{roy13}. We now exploit the large SHS produced in our system  to induce spinning motion in individual trapped peapod-shaped soft oxometalate \cite{sou11} or quartz microparticles (in accordance with the scheme displayed in Fig.~\ref{elliptfig}(f)). The choice of asymmetric particles was driven by the fact that such particles exhibit form birefringence \cite{bohren}. In the presence of elliptically polarized light, birefringent particles experience two types of torque -- an alignment torque that aligns the fast axis of the particle along the electric field, and a spinning torque that is proportional to the degree of ellipticity, light intensity, and thickness of the particle \cite{fri98}. For purely circularly polarized light, the alignment torque vanishes and the particle exhibits pure rotation.

We show the following distinct effects: (a){\it Rotation (spinning) of single particles using a linearly polarized Gaussian beam} and (b){\it Control of the rotation using a second beam}. We show rotation of a single peapod (clockwise) and quartz particle (anticlockwise), respectively, in Fig.~\ref{exptdata}(b) and (c) (see also video 1, 2, and 3 in Supplementary Information B which show clockwise and anticlockwise rotation of a peapod, and anticlockwise rotation of a quartz particle, respectively). Fig.~\ref{exptdata}(d) shows the rotation speed of a representative peapod (crosses) and quartz particle (open circles) as a function of incident laser power. The quartz particles typically rotate faster (few Hz) since they are on the average bigger and thicker than the peapods (sub-Hz rotation rate). Control of rotation is demonstrated by introducing a second beam of the same polarization (see Fig.~\ref{exptdata}(e)), which can either stop the rotation or else change the sense of rotation depending on its intensity with respect to the first beam. Videos 3 and 4 in Supplementary Information B demonstrate stopping and changing the direction of rotation of single quartz micro-particles by the control beam, respectively. 

\section{Discussion}
Spinning of particles about their axes has also been reported in literature to be a consequence of the unbalanced torque arising due to the asymmetry in scattering from irregularly shaped particles \cite{niem09}. For these cases, the asymmetry in scattering needs to be strong enough to exceed the viscous drag force of the fluid medium which surrounds the particles. However, we have verified by simulation that such strong asymmetry does not arise for the ellipsoidal shaped particles that we observe rotating (Fig.~\ref{exptdata}(b) and (c)) in our experiments. Indeed, for scattering to cause rotations, the particles need to have large extrusions on them so as to bring about the strong azimuthal scattering asymmmetry required to produce the necessary torque. Thus, we can safely conclude that the observed rotations on the particles are a consequence of the enhanced SOI. 

We believe that our findings on the fascinating manifestations of the enhanced SOI due to the propagation of a tightly focused beam in a stratified medium could lead to further explorations on these systems. While we have experimentally demonstrated interesting applications in controlled rotation and transportation of micro-particles, several other applications can also be envisaged which may open new directions of research in optical tweezers. For example,  micro-particles can be trapped in multiple radial rings (Fig.~\ref{diatt}(f)) similar to holographic tweezers, but by using a single Gaussian beam. The presence of annular regions of high ellipticity could also produce multi-particle trapping in combination with rotations in both directions using a single Gaussian beam - a configuration that could enable the study of optical binding as well as exchange of angular momentum between birefringent particles.

In conclusion, we investigate the many interesting manifestations of SOI due to a tightly focused Gaussian beam propagating in a stratified optical trap.  We study the effects of both spin redirectional topological phase as well as polarization dependent trajectory (SHE). The effect of the topological phase is shown to introduce a large anisotropic diattenuation that modifies the radial intensity distribution near the focal plane. For a cover slip that is chosen to be a polarizer for certain spatial harmonics, the structure of a Gaussian beam is modified to the extent of the formation of discrete off-axis intensity lobes around the center at the focus. For  partially polarizing cover slips, a high RI value causes the formation of intensity lobes in the background of a continuous ring that can be used to transport particles \cite{roy13}, as well as multiple intensity rings that can support the trapping of particles similar to holographic tweezers in the radial direction. The enhanced SOI also causes a large SHE to break down the incident linearly polarized Gaussian beam into components of large degree of opposite circular polarization that are spatially separated due to a large SHS near the trap focal plane. Asymmetric particles trapped at the epicentres of such regions can be rotated (spun) with full control on their rotational degree of freedom.

\section{Acknowledgements}
The authors would like to acknowledge Arijit Haldar for the EM field analysis, and Atharva Sahasrabuddhe and Bibudha Parasar for help in preparing the peapod and quartz samples. This work was supported by the Indian Institute of Science Education and Research, Kolkata, an autonomous research and teaching institute funded by the Ministry of Human Resource Development, Govt. of India. SR thanks DST Fast track and BRNS DAE grants.
\section{Methods}
The experiments are performed for asymmetric particles that have form birefringence by virtue of their elongated structure. We use two types of particles: peapod shaped soft oxometalates (Ammonium phosphomolybdate, $(NH_4)_3[PMo_{12}O_{40}]$)) \cite{sou11} of average dimension $1.5 \times 0.5~\mu$m and quartz microparticles of generally arbitrary shape (obtained by crushing a large quartz crystal) but of dimensions between 1-3 $\mu$m. The optical tweezers is developed around a Zeiss Axiovert.A1 inverted microscope with an oil immersion 100X 1.41 NA objective focusing the trapping linearly polarized laser at 1064 nm into the sample chamber. The polarization of the beam can be controlled by a half-wave retarder placed at the input of the trap. Around 25 $\mu$l of a dilute aqueous solution of the microparticles is placed in the sample chamber consisting of a cover slip and a standard microscope slide of thickness 1 mm (bottom surface) resulting in a sample depth of about 30-40 $\mu$m. As mentioned earlier, our cover slip is not index-matched with the objective immersion oil, which is very different from standard optical tweezers configurations and results in a stratified medium in the forward direction which enhances the SOI of the tightly focused trapping light. 

For some of our experiments on controlling the rotation of trapped particles, we use a non-polarizing beam splitter cube at the input of the microscope to generate two independent trapping beams to form two traps that are adjacent to each other. The power levels of the two trapping beams are controlled by placing neutral density filters in the path of the beam reflected from the cube, as shown in Fig.~\ref{exptdata}(a). To visualize the particles, an IR filter is required in front of the camera to avoid saturating it, which is removed in image 4 in Fig.~\ref{exptdata}(a) where the a ring-like intensity structure is evident with the rotating particle located on the ring at the left of the central maxima. This signifies that it is indeed sampling a region of high ellipticity that causes it to spin (see Video 1 of Supplementary Information B). Anti-clockwise rotation of a peapod is shown in Video 2 of Supplementary Information B, with the peapod being now located at the right corner of the intensity ring (obvious from the fact that the central maxima now appears at the left corner of the video). The rotation frequencies of single trapped peapods and quartz particles are obtained by detecting the back-scattered intensity from a weak detection laser at 670 nm that is incident on the trapped particle. The output signal is modulated at the frequency of rotation. 

\providecommand{\noopsort}[1]{}\providecommand{\singleletter}[1]{#1}

\end{document}


\title{Supplementary Information A}
\author{Basudev Roy$^1$}
\author{Nirmalya Ghosh$^1$} \email{nghosh@iiserkol.ac.in}
\author{S. Dutta Gupta$^2$}
\author{Soumyajit Roy$^3$}
\author{Ayan Banerjee$^1$} \email{ayan@iiserkol.ac.in}
\affiliation{$^1${Department of Physical Sciences, IISER-Kolkata, Mohanpur 741252, India}}
\affiliation{$^2$School of Physics, University of Hyderabad, Hyderabad 500046, India}
\affiliation{$^3${EFAML, Materials Science Centre, Department of Chemical Sciences, IISER-Kolkata, Mohanpur 741252, India}}
\maketitle
\noindent 
\section{\label{emfield}Theoretical calculations}
\subsection{Expressions for transfer function $A$ and $I_0, I_1$, and $I_2$} We use the Debye-Wolf theory  \cite{born, rich59, blio11} to determine the exact nature of the electric field distribution due to tight focusing of a linearly polarized Gaussian beam in a stratified medium. In this theory, an incident collimated Gaussian beam is decomposed into a superposition of plane waves having an infinite number of spatial harmonics. A longitudinal component emergences in the total intensity of the light after tight focusing, and the resulting field amplitude can be related to the incident field  by the action of a transfer function that can be written as $A = R_z(-\phi)R_y(\theta)R_z(\phi)$, where $R_i(\alpha),~i=x, y, z$ represents the SO(3) rotation matrix around the $i$ axis by angle $\alpha$. Here, $\phi$ could be understood as the azimuthal angle, while $\theta$ is the polar angle defined with respect to $x$ and $z$ axis of the laboratory frame respectively. The stratified medium introduces a polarization dependence of the field propagating in the medium, so that $A$ needs to incorporate  $T_s~(R_s)$ and $T_p~(R_p)$ -- the Fresnel transmission (reflection) coefficients (generally complex) which include the multiple interface contributions for $s$ and $p$ polarizations respectively. Then, the resultant field amplitude $\vec{E}_{res}(\theta,\phi)$ can be written in terms of the incident amplitude $\vec{E}_{inc}(\theta,\phi)$ as
\begin{equation}
\label{fieldinout}
\vec{E}_{res}(\theta,\phi)=A\vec{E}_{inc}(\theta,\phi),
\end{equation}
where the transfer function $A$ is given by 
\begin{widetext}
\begin{eqnarray}
\label{transfermatrix}
A_{1,j}^t &=  R_z(-\phi)R_y(\theta)TR_z(\phi) =  
&= \left[
\begin{array}{ccc}
\ a - b\cos 2\phi  &\ -b\sin 2\phi &\ c\cos\phi\ \\
\  -b\sin 2\phi &\ a + b\cos 2\phi &\ c\sin\phi\ \\
\ -c\cos\phi &\ -c\sin\phi &\ a-b\ \\
\end{array}
\right ].
\end{eqnarray}
\end{widetext} 

In the case of forward-propagation, the coefficients of $A$ are given by $a = \frac{1}{2} \left (T_s+T_p \cos\theta \right )$, $b = \frac{1}{2}\left (T_s-T_p \cos\theta \right )$, and $c =  T_p\sin\theta$, while for the backward propagating waves the coefficients in eq.~\ref{transfermatrix} would be modified with $\theta$ replaced by $\pi - \theta$, and the Fresnel reflection coefficients $R_s$ and $R_p$ being used instead of the transmission ones. While in general, $\vec{E}_{res}(\theta,\phi)$ would be a superposition of forward and backward propagating waves in the stratified medium, the dominant contribution would come from the forward propagating  waves.  The final field is obtained by integrating Eq.~\ref{fieldinout} over $\theta$ and $\phi$, so that
\begin{eqnarray}\label{polarint}
\vec{E}(\rho,\psi,z)&=&  i\frac{kfe^{-ikf}}{2\pi}\int_0^{\theta_{max}}\int_0^{2\pi}
\vec{E}_{res}(\theta,\phi)e^{ikz\cos\theta}\nonumber\\& \times & e^{ik\rho\sin\theta\cos(\phi-\psi)}
sin(\theta)\>d\theta d\phi,
\end{eqnarray}
where $r = f$ -- the focal length of the lens. The limit for the $\theta$ integral is set by the numerical aperture of the microscope objective. Now, for incident $x$-polarized light represented by a Jones vector  $\left[\ 1\ 0\ 0\ \right]^T$), the electric field can be written from eq.~\ref{polarint} as
\begin{eqnarray}\label{fieldout}
\left[
\begin{array}{c}
 {E_x}\\
 {E_y}\\
 {E_z}\\
\end{array}
\right ] &=& C \left[
\begin{array}{lll}
 I_0 + I_2\cos 2\psi  & I_2\sin 2\psi & 2i I_1\cos\psi \\
 I_2\sin 2\psi & I_0 - I_2\cos 2\psi & 2iI_1\sin\psi \\
 -2iI_1\cos\psi & -2iI_1\sin\psi & I_0+I_2 \\
\end{array}
\right ]\nonumber\\  && \times \left[
\begin{array}{c}
 {1}\\
 {0}\\
 {0}\\
\end{array}
\right ]
 =  C
\left[
\begin{array}{c}
 {I_0+I_2\cos 2\psi }\\
 {I_2\sin 2\psi }\\
 {-i2I_1 \cos \psi }\\
\end{array}
\right ].
\end{eqnarray}
The transmitted and reflected components of $I_0(\rho),~I_1(\rho)$ and $I_2(\rho)$ would be (suffixes $t$ and $r$ imply transmitted and reflected respectively) given by
\begin{widetext}
\begin{eqnarray}
\label{transmisseqn}
I^t_0(\rho)=\int_0\limits^{min(\theta_{max},\theta_c)}E_{inc}(\theta)
\sqrt{\cos\theta}(T^{(1,j)}_s+T^{(1,j)}_p\cos\theta_j)J_0(k_1\rho\sin\theta)e^{ik_jz\cos\theta_j}sin(\theta)\>d\theta, \nonumber
\end{eqnarray}
\begin{eqnarray}
I^t_1(\rho)=\int_0\limits^{min(\theta_{max},\theta_c)}E_{inc}(\theta)
\sqrt{\cos\theta}T^{(1,j)}_p\sin\theta_j
J_1(k_1\rho\sin\theta)e^{ik_jz\cos\theta_j}\sin\theta\>d\theta, \nonumber
\end{eqnarray}
\begin{eqnarray}
I^t_2(\rho)=\int_0\limits^{min(\theta_{max},\theta_c)}E_{inc}
(\theta)\sqrt{\cos\theta}(T^{(1,j)}_s-T^{(1,j)}_p\cos\theta_j) J_2(k_1\rho\sin\theta)e^{ik_jz\cos\theta_j}\sin\theta\>d\theta,
\end{eqnarray}
\end{widetext}
and 
\begin{widetext}
\begin{eqnarray}
I^r_0(\rho)=\int_0\limits^{min(\theta_{max},\theta_c)}E_{inc}(\theta)\sqrt{\cos\theta}
(R^{(1,j)}_s-R^{(1,j)}_p\cos\theta_j)J_0(k_1\rho\sin\theta)e^{-ik_jz\cos\theta_j}\sin\theta\>d\theta, \nonumber
\label{reflecteqn}
\end{eqnarray}
\begin{eqnarray}
I^r_1(\rho)=\int_0\limits^{min(\theta_{max},\theta_c)}E_{inc}(\theta)\sqrt{\cos\theta}
R^{(1,j)}_p\sin\theta_k J_1(k_1\rho\sin\theta)e^{-ik_jz\cos\theta_j}\sin\theta_1\>d\theta, \nonumber
\end{eqnarray}
\begin{eqnarray}
I^r_2(\rho)=\int_0^{min(\theta_{max},\theta_c)}E_{inc}(\theta)\sqrt{\cos\theta}
(R^{(1,j)}_s+R^{(1,j)}_p\cos\theta_j) J_2(k_1\rho\sin\theta)e^{-ik_jz \cos\theta_j}\sin\theta\>d\theta,
\end{eqnarray}
\end{widetext}
where the $\phi$ integrals have been carried out and are related to Bessel functions $J_n$.
\section{\label{3dstokes} Three dimensional analysis of polarization}
\subsection{Three dimensional stokes vector components and degree of circular polarization}
The SHE owes its origin to the longitudinal components of the electric field which results in transverse energy flow, given by the transverse components of the Poynting vector. Thus, in order to calculate the SHE, a 3d treatment of polarization is warranted. We therefore proceed to build the 3d coherency matrix of the system that is in general given by \cite{set02} 
\begin{eqnarray}\label{coherencymat}
\Phi &=& C \left[
\begin{array}{lll}
\langle E_x^*E_x 
\rangle & 
\langle E_x^*E_y \rangle & 
\langle E_x^*E_z\rangle \\
\langle E_y^*E_x\rangle & 
\langle E_y^*E_y\rangle& 
\langle E_y^*E_z\rangle\\
\langle E_z^*E_x\rangle & 
\langle E_z^*E_y\rangle & 
\langle E_z^*E_z\rangle \\
\end{array}
\right ]
\end{eqnarray}
The polarization states of light are typically represented by Stokes vectors, with $4\times 1$ column vectors used for 2-d polarization states. In the case of  3-d polarization, we need to define $9\times 1$ Stokes vectors $\Lambda_{i, i=1, 9}$ to include the longitudinal polarization component as well. 
The 3-dimensional stokes vector components are obtained by writing $\phi = \dfrac{1}{3}\Sigma{_{i=0}^8}~\lambda_i S_i$, where $\lambda_0$ is the identity matrix, and $\lambda_i(i=1,8)$ are the Gell-man matrices \cite{set02} which are Hermitian, trace-less, and linearly independent. Then, we have 
\begin{equation}
\label{3dstokes0}
S_{0}=E_x E^{*}_x+E_y E^{*}_y+E_z E^{*}_z
\end{equation}
\begin{equation}
\label{3dstokes1}
S_{1}=(3/2)(E_x E^{*}_y+E_y E^{*}_x)
\end{equation}
\begin{equation}
\label{3dstokes2}
S_{2}=(3i/2)(E_x E^{*}_y-E_y E^{*}_x)
\end{equation}
\begin{equation}
\label{3dstokes3}
S_{3}=(3/2)(E_x E^{*}_x-E_y E^{*}_y)
\end{equation}
\begin{equation}
\label{3dstokes4}
S_{4}=(3/2)(E_x E^{*}_z+E_z E^{*}_x)
\end{equation}
\begin{equation}
\label{3dstokes5}
S_{5}=(3i/2)(E_x E^{*}_z-E_z E^{*}_x)
\end{equation}
\begin{equation}
\label{3dstokes6}
S_{6}=(3/2)(E_y E^{*}_z+E_z E^{*}_y)
\end{equation}
\begin{equation}
\label{3dstokes7}
S_{7}=(3i/2)(E_y E^{*}_z-E_z E^{*}_y)
\end{equation}
\begin{equation}
\label{3dstokes8}
S_{8}=(\sqrt(3)/2)(E_x E^{*}_x+E_y E^{*}_y-2 E_z E^{*}_z)
\end{equation}

Among these components, the $S_{0}$ indicates total intensity, the $S_{2}$, $S_{5}$ and $S_{7}$ indicate the degree of circular polarization in the XY, ZX and YZ planes respectively. 

\subsection{\label{circfield}Components of the electric and magnetic field and the Poynting vector for circularly polarized light}
The corresponding components of the electric and magnetic field for circularly polarized light are given by the following expressions
\begin{equation}
\label{cEx}
E_{x}=(I_0+I_2 cos(2\psi))+i(I_2 sin(2(\pi/2-\psi)))
\end{equation}
\begin{equation}
\label{cEy}
E_{y}=(I_2 sin(2\psi))+i(I_0 + I_2 cos(2(\pi/2-\psi)))
\end{equation}
\begin{equation}
\label{cEz}
E_{z}=-2i I_1 cos(\psi)-2 I_1 cos((\pi/2-\psi))
\end{equation}
\begin{equation}
\label{cBx}
B_{x}=-i(I_0+I_2 cos(2\psi))+(I_2 sin(2(\pi/2-\psi)))
\end{equation}
\begin{equation}
\label{cBy}
B_{y}=-i(I_2 sin(2\psi))+i(I_0 + I_2 cos(2(\pi/2-\psi)))
\end{equation}
\begin{equation}
\label{cBz}
B_{z}=-2 I_1 cos(\psi)-2i I_1 cos((\pi/2-\psi))
\end{equation}

These components can be used to determine the components of the Poynting vector as 
\begin{equation}
\label{cSx}
P_{x}=E_y B^{*}_z-E_z B^{*}_y
\end{equation}
\begin{equation}
\label{cSx}
P_{y}=E_z B^{*}_x-E_x B^{*}_z
\end{equation}
\begin{equation}
\label{cSx}
P_{z}=E_x B^{*}_y-E_y B^{*}_x
\end{equation}
The extent of the Spin-Hall shift for circularly polarized light is given by the following.
\begin{equation}
\Delta y = \pm \dfrac{\Sigma y (P_x + P_y)}{\Sigma P_i},
\label{shs}
\end{equation}
with the positive and negative signs corresponding to left and right circularly polarized light respectively, and $i=x,y,z$. Note that for input linearly polarized light, the SHS would be $2\Delta y$.
\providecommand{\noopsort}[1]{}\providecommand{\singleletter}[1]{#1}